\documentclass[preprint,12pt]{elsarticle}

\usepackage{amssymb}
\usepackage{color}
\journal{Nuclear Inst. and Methods in Physics Research, A}

\begin{document}

\begin{frontmatter}



\title{Optimising a Muon Spectrometer for Measurements at the ISIS Pulsed Muon Source}


 \author[label1,label2]{S.R. Giblin}
 \author[label1]{S.P. Cottrell}
 \author[label1]{P.J.C. King}
 \author[label1]{S. Tomlinson}
 \author[label1]{S.J.S. Jago}
 \author[label1]{L.J. Randall}
 \author[label1]{M.J. Roberts}
 \author[label1]{J. Norris}
 \author[label1]{S. Howarth}
 \author[label1]{Q.B. Mutamba}
 \author[label1]{N.J. Rhodes}
 \author[label1]{F. Akeroyd}
\address[label1]{ ISIS Facility, Rutherford Appleton Laboratory,
Chilton, Didcot, Oxon, OX11 0QX, UK}
\address[label2]{School of Physics and Astronomy, Cardiff University, Cardiff, CF24 3AA, UK.}

\begin{abstract}
This work describes the development of a state-of-the-art muon spectrometer for the ISIS pulsed muon source. Conceived as a major upgrade of the highly successful EMU instrument, emphasis has been placed on making effective use of the enhanced flux now available at the ISIS source. This has been achieved both through the development of a highly segmented detector array and enhanced data acquisition electronics. The pulsed nature of the ISIS beam is particularly suited to the development of novel experiments involving external stimuli, and therefore the ability to sequence external equipment has been added to the acquisition system. Finally, the opportunity has also been taken to improve both the magnetic field and temperature range provided by the spectrometer, to better equip the instrument for running the future ISIS user programme.
\end{abstract}

\begin{keyword} 
muon spectrometer
\sep EMU
\sep ISIS pulsed muon facility
\end{keyword}

\end{frontmatter}

% \linenumbers

\section{Introduction.}
\label{intro}
The muon spin relaxation ($\mu$SR) technique involves the implantation of intrinsically 100$\%$ spin polarised positively charged muons ($\mu ^+$) to probe condensed matter systems. Owing to its spin (I=1/2), the muon couples to and provides information about the local magnetic environment, while the mass ($m_\mu \approx m_p/9$) enables the muon to mimic a very light hydrogen isotope. In certain materials muonium may be formed as a tightly bound $\mu ^+ e^-$ species (analogous to the hydrogen atom) to provide a powerful chemical probe. These unique properties have led to numerous and diverse applications of the $\mu$SR technique~\cite{SJB,ms,yau} including the study of magnetism and associated spin dynamics, superconductivity, the hydrogen environment and diffusion, and the chemistry associated with radical states. The scope of the technique, however, presents a challenge to the facility scientist, who is required to develop instrumentation that is capable of satisfying the often contradictory demands of these various user communities studying all three phases of matter. For example, the spectrometer is expected to be easily configurable to run from 20mK to 1400K, and then within a few hours be adapted to run a gas cell with a triggered radio frequency (RF) excitation.

The EMU $\mu$SR spectrometer was commissioned in 1993 as part of a major upgrade of the European Muon Facility to deliver single muon pulses simultaneously to three experimental areas~\cite{gordon} through the use of a fast electric field (EF) kicker. At ISIS, beams of $\sim$4 MeV/c surface muons are produced in $\sim$80 ns, full width half maximum (FWHM) wide pulses with a repetition rate of 40 Hz, with each muon pulse defined as one ISIS frame. Thus, muons are produced every 20 ms (but one pulse in five is diverted to the ISIS second target station) with data acquisition continuing for ~32 $\mu$s following the muon arrival. Muons are guided to the spectrometer by a series of dipole and quadrupole magnets, forming an uncollimated muon spot of $\sim$25x10 mm$^2$ at the sample position. Muons have a lifetime of $\sim$2.2 $ \mu$s and are typically detected by measurement of their decay products. The instrument detector array consists of rings of symmetric scintillation counters positioned upstream (forward) and downstream (backward) relative to the sample position to follow the time evolution of the decay positron distribution. The original EMU array was segmented into 32 elements to permit measurement without distortion caused by the high instantaneous rates that are inherent to a pulsed source following muon implantation. The number of elements chosen was well matched to then typical ISIS beam currents, with each muon pulse containing $\sim$300 muons, corresponding to an average of 3 hits per ISIS frame per detector over the entire array. Finally, the solid angle of the array was designed to compensate for the 50$\%$ reduction in muon flux arising from the spatial splitting of the beam by the  kicker. The instrument incorporated a vacuum chamber to minimise the number of beam windows and associated scatter when using cryogenic sample environment, while enabling the use of suspended samples for small sample measurements~\cite{gordon2,lynch}. Resistive coils provided 0.4 T and 0.01 T fields parallel and perpendicular to the incident muon beam respectively, while three orthogonal low field ($\sim$0.2 mT) air-cooled coil pairs were incorporated to enable environmental fields to be nulled to create a true zero field condition at the sample position.

The spectrometer operated as part of the muon facility at ISIS for about 17 years or approximately 3000 beam days, delivering over 600 separate user experiments over its lifetime. Over this period, however, both the development of the ISIS source and an increase in the thickness of the muon production target combined to increase muon production by a factor of approximately six. To make effective use of this enhanced flux the design of both the detector array and counting electronics needed to be reconsidered, with this providing a valuable opportunity to revisit other aspects of the instrument performance. Recently there has been a significant growth in the number of groups seeking to carry out experiments that involve external stimuli and exploit the pulsed nature of the ISIS beam. In this type of experiment excitations of either the muon or the sample, depending on the specific measurement, are synchronised with muon implantation. This includes the development of RF and EF techniques, together with novel applications of pump-probe laser excitation. To facilitate this work, the opportunity was taken to provide the acquisition system with a flexible method of sequencing the switching of multiple pieces of external equipment, collecting data for each separate experimental state.

\section{Designing a state-of-the-art Spectrometer}
\subsection{Instrument detector array}
A number of factors needed to be considered when designing the array for the upgraded instrument:
\begin{enumerate}[i]
\item{Increased detector segmentation is clearly essential for measuring high data rates without the distortion associated with signal pile-up (when one or more positrons strike the same scintillating element within the deadtime of the first). However, increasing the number of scintillation counters carries both a financial and engineering cost, with the latter relating to the difficulty of mounting multiple photomultiplier tubes and light guides and the problem of the scintillator wrapping reducing the available active area. While the application of a multichannel detector~\cite{riken} or the development of a Geiger-mode avalanche photodiode based detector array~\cite{psi} may eventually overcome these difficulties, it was preferred that the new array be based on established technologies to facilitate rapid development. The final segmentation for the detector array was derived by considering the typical sample size ($\sim$24 mm diameter) and the appropriate setting of the collimator slits incorporated in the beamline ($\sim$15 mm), chosen to avoid an excessive signal from around the sample. From a calibration of beam intensity, for efficient use of available beam it was concluded that an array capable of handling data rates of up to $\sim$500 detected events per ISIS frame was required. With a typical deadtime for each segment of $\sim$10 ns, an average detection rate of less than five hits per ISIS frame per detector element was required to avoid measurable distortion for typical run statistics, and therefore a detector segmentation of $\sim$100 elements was appropriate for the design. In fact, a configuration using 96 elements was selected after considering engineering constraints, with novel data acquisition electronics being developed to permit higher counting rates through effective deadtime correction.}\\

\item{While there was a desire to maximise the solid angle coverage of the new detector array, its design was constrained by the overall configuration of the instrument including the beam path and the need to ensure enough space was left to mount the sample environment in the spectrometer. (Figure \ref{detector}a). In practice the solid angle coverage for the new detector array is similar to that used on the original instrument ($\sim$2.2 $ \pi$ sr).}\\

\item{A growing number of experiments measure small samples ($<$10 mm diameter) where the signal from around the sample would dominate given conventional mounting on a large silver plate. To overcome this problem the technique of suspending the sample in the beam is typically employed~\cite{lynch}, where muons not implanted in the sample continue in an evacuated flight tube, are removed from the sample position and no longer contribute to the measured signal. Unfortunately, because data rates from the sample are low, the signals measured using this technique are particularly susceptible to distortion from counts originating outside the sample. The origin of these stray counts is currently unknown, although previous work~\cite{lynch} suggests they originate from muons scattered by the beamline window into the walls of the cryostat and instrument vacuum vessel. To minimise their effect on the data both the forward and backward detector banks were subdivided into three stepped rings, each ring containing 16 scintillating elements.
The total solid angle of the new detector array is identical to the old array; however, the addition of a third ring offers greater flexibility for removing contaminated detectors from the data analysis whilst maintaining the maximum possible count rate. A cross section through the array showing the general arrangement of the scintillating elements is shown in Figure \ref{detector}b. During tests, this geometry appeared to maximise the possible data rates as two of the three rings were generally found to be completely unaffected by distortion in the data - presumably the individual rings make a different solid angle to the source of the stray counts. The design therefore gives great flexibility for selecting detector rings to provide clean data during analysis.}\\

\begin{figure}[t]
\centering
\includegraphics[width=7cm]{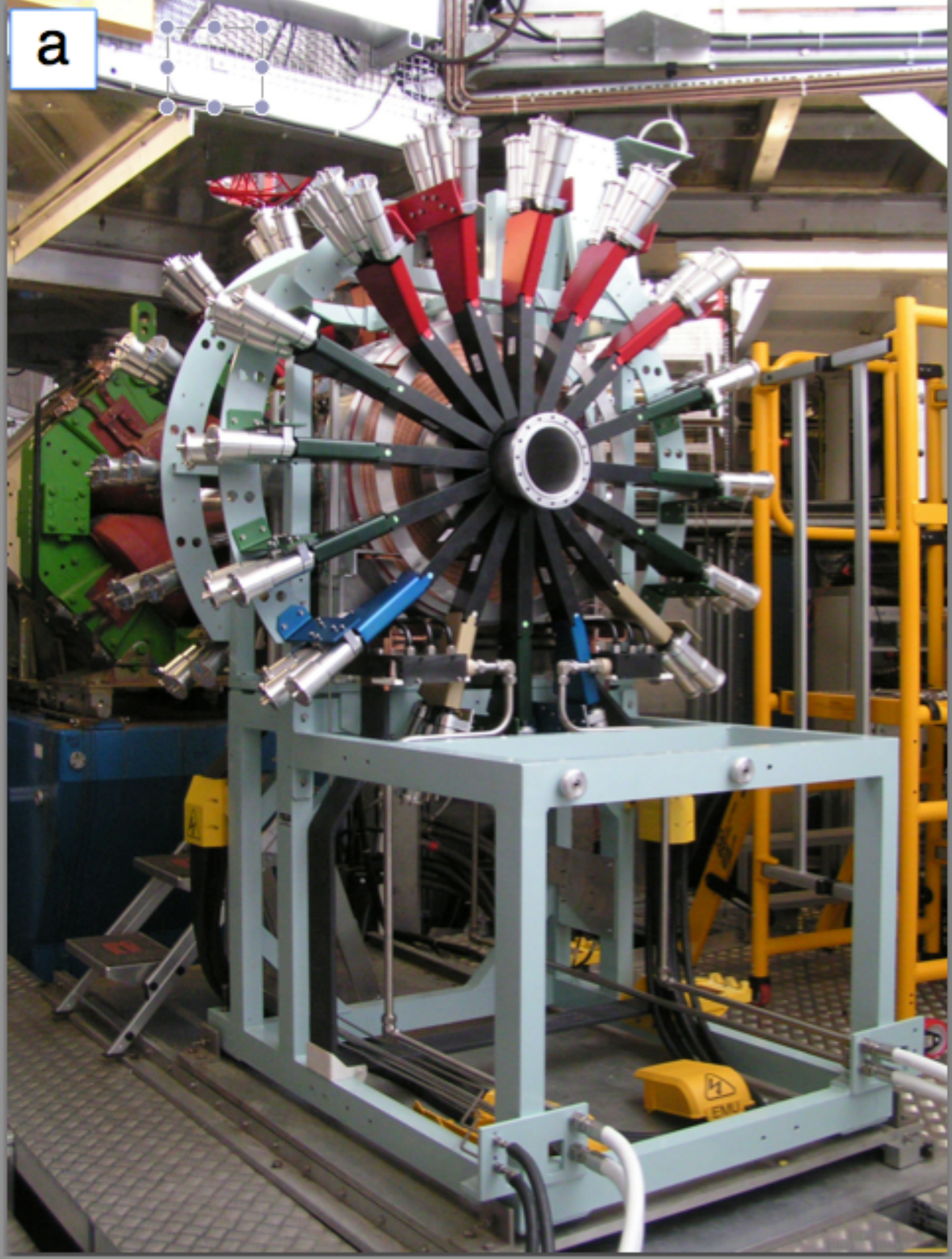}
\includegraphics[width=12cm]{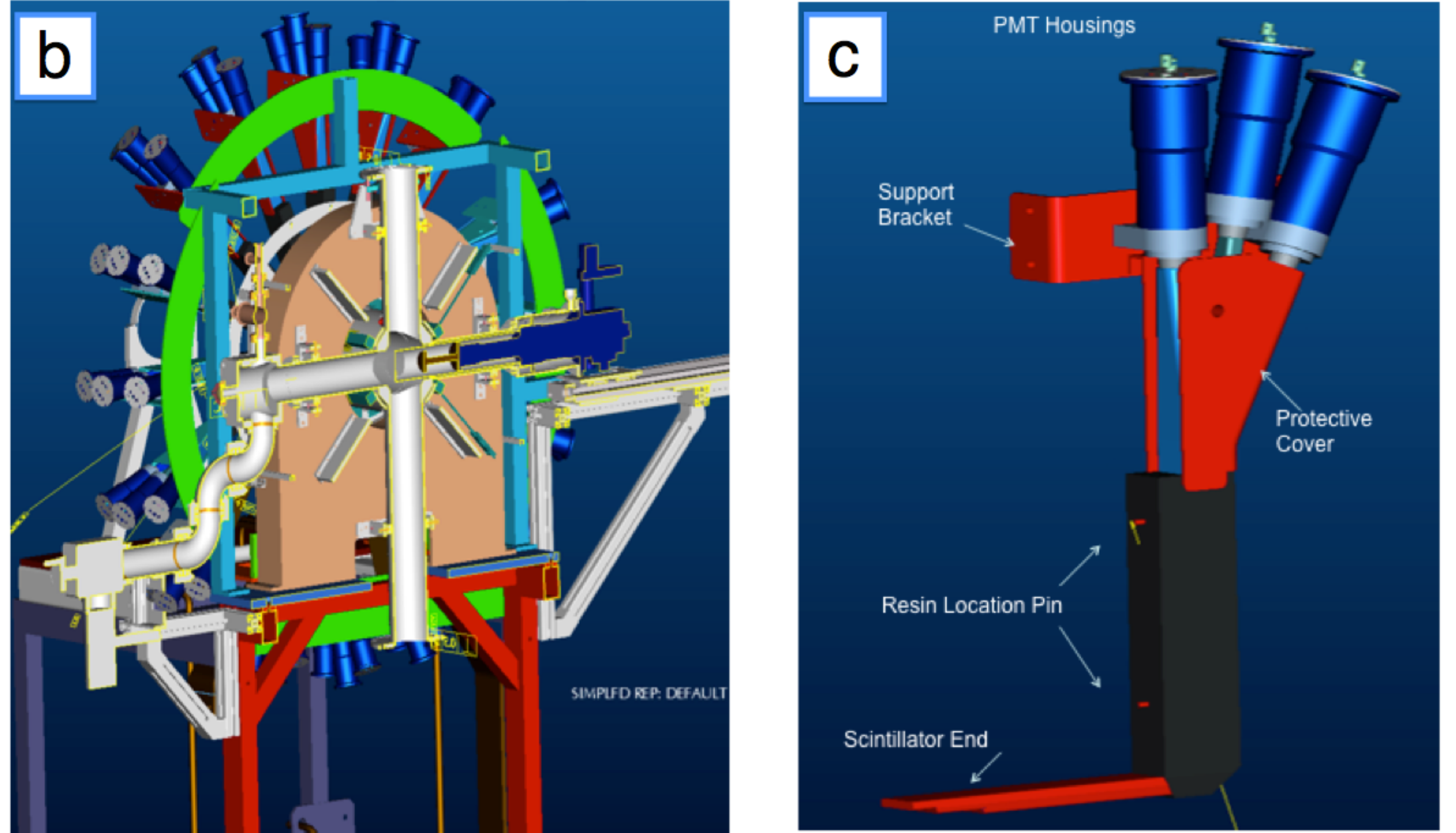}
\caption{a) Complete instrument, where the beam centre is  at a height of 1855mm. b) Cross section of the instrument showing the vacuum cruciform (with sample blade inserted on a cryostat from the right) magnet and one side of the detector arrangement consisting of 16 units each with 3 PMT, making a final total of 96 detectors. c) The final detector unit used, there are three scintillator ends (creating the inner, middle and outer rings) of different lengths to ensure nominally the same solid angle.}\label{detector}
\end{figure}

\item{Robust construction of the detector elements was essential, and to this end comparatively short light guides ($\sim$1 m) were designed with corresponding stepped elements formed into a single unit by encapsulating the guides in resin. The final units, shown in Figure \ref{detector}c, could then be easily located on the spectrometer to build up the full detector array. Magnetic field tolerant photomultiplier tubes (PMT) made by Hamamatsu\cite{ham} (R5505, running with positive polarity) were used owing to their proximity  to the centre of the 0.5 T magnet. 
The tubes were mounted unscreened with their axes approximately perpendicular to the ~0.05 T stray field. In this configuration, measurements of the PMT pulse amplitudes confirmed their gain was invariant with magnetic field, essential to avoid artefacts occurring in the data during magnetic field sweeps.}
\end{enumerate}

\subsection{Data acquisition elements}
Analogue signals from the photomultiplier tubes are conditioned using CAEN\cite{caen} V895B leading edge discriminator modules, the units have beencaen modified by the manufacturer for improved input frequency (250 MHz) and pulse pair resolution (4 ns). PMT voltages were optimised with reference to the pulse height spectrum, and set such that a discriminator threshold value of 75 mV clearly separated the signal peak associated with positron decays from noise (a valid event). Signal output from each discriminator is passed to a time to digital converter (TDC) 
where positron decay events are time stamped. This allows a histogram of valid events to be formed over a defined time window of $\sim$32 $\mu$s relative to a common start signal (obtained from a Cherenkov detector located in the muon beamline).

Until recently, the muon instrument suite at ISIS used a (now obsolete) LeCroy\cite{lecroy}  MTD133B 8 channel multi-hit TDC, these CMOS integrated circuits being built into a custom VXI-based acquisition system in-house. While functional, significant problems were encountered at higher event rates with distortion of the early time data owing to the limited 16 event LIFO (Last In First Out) buffer contained within these devices.  An analysis frequently used for characterizing and correcting the dead time associated with elements in a $\mu$SR detector is the non-extendable model discussed by Leo~\cite{leo}, where events are simply unrecorded during the dead period without extending the time the detector is unresponsive. However, using this type of analysis it proved impossible to correct for the effect of buffer overrun in the LeCroy\cite{lecroy}  TDC. As a result, experiments that might naturally have run at higher rates (those with large samples or having a particular interest in results at long-times, for example) and relied on correcting early time data for curve fitting were frequently compromised. As part of the upgrade programme there was a clear desire to improve the performance of the TDC at high data rates, both by enhancing the general specification and ensuring that any deadtime contribution was amenable to correction by standard methods.

\begin{figure}[t]
\centering
\includegraphics[width=18cm, angle=90]{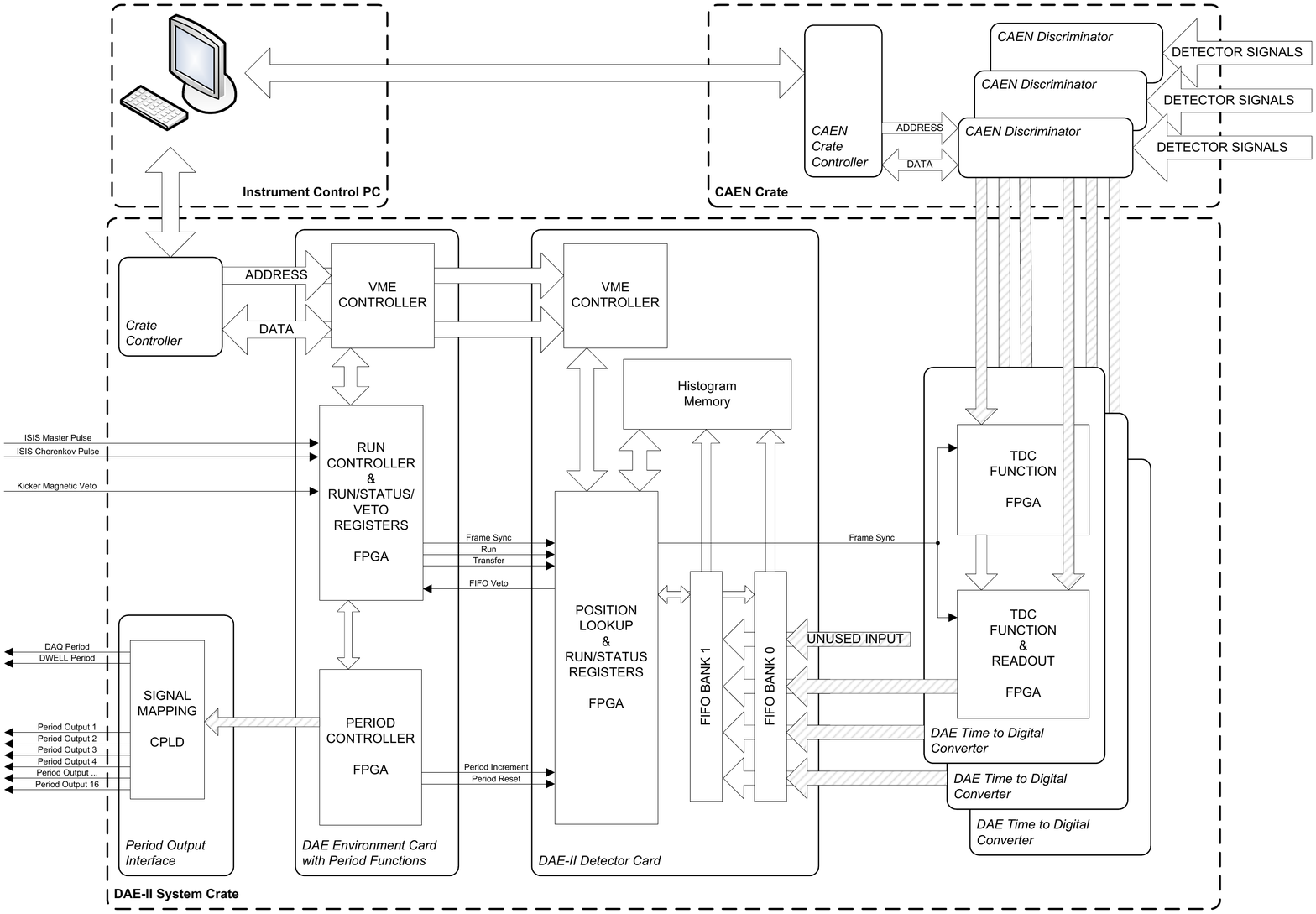}
\caption{A block diagram for the data acquisition electronics as described in the text providing a fully configurable system to control novel experiments.}\label{electronics}
\end{figure}

Without a satisfactory commercial alternative, it was decided to implement a replacement 16 channel TDC using an Altera\cite{altera} Stratix III (EP3SL50F484C2) field programmable gate array (FPGA). Two of these devices were combined on a single VXI card to provide a high density 32 channel TDC acquisition module. The device is designed to collect multiple events over a time window of $\sim$32 $\mu$s following a common start signal. Time bins may be set to between 2 ns and 16 ns according to the type of measurement being made. Internally, while the device is running it creates a record of time stamped events relative to the common start signal. After capture is complete the events are moved to a separate detector card where they are sorted and added to the histograms. The final histograms are uploaded to the control computer at the end of the run, although a ``snap shot'' may be taken at any time to monitor the evolving measurement. A block diagram of the complete data acquisition system is shown in Figure \ref{electronics}. Crucially, the ultimate limit to the pulse pair resolution is set only by the bin width, and the number of events that can be captured within an ISIS frame constrained only by the transfer speed between the TDC and the downstream electronics ($\sim$512 events/channel assuming a 20 ms window between ISIS frames). Both parameters are a significant improvement compared to the MTD133B, where the pulse pair resolution is 10 ns and the number of buffered events is limited to 16.

\subsection{Data ``Period'' switching}
Experiments frequently require the muon response to be compared when different external stimuli (e.g. RF, laser illumination, EF, etc) are applied to the sample. To simplify comparison while removing systematic errors, data acquisition is divided into a number of acquisiton ``Periods''; for each ``Period'' external equipment is switched into a unique state and data stored in a separate histogram memory. Each ``Period'' is measured for a specific number of ISIS frames and a pre-defined sequence of ``Periods'' repeats continuously until acquisition is stopped. The acquisition electronics ensures histograms are acquired only from complete ``Period'' sequences.

To provide this functionality, a ``Period'' card was designed with a pair of Altera\cite{altera} Apex 20KE FPGAs (EP20K400EFC672-2X) operating in parallel, one running the standard DAQ functions, the other ``Period'' functions. Specifically, each ``Period'' can record between 1 to 64x10$^3$ ISIS frames and may be defined as either a DAQ ``Period'', where data is collected into an area of histogram memory corresponding to the ``Period'' number in sequence, or a Dwell ``Period'' where ISIS frames are counted, but data is vetoed by the acquisition electronics. Dwell ``Periods'' can therefore be used to allow time for the external stimuli to stabilise prior to the acquisition of data in a DAQ ``Period''.  The total number of ``Periods'' programmable in each sequence is 16384, although for most practical purposes only a small number will be required as the sequence is repeated until the necessary statistics has been acquired. The ``Period'' controller synchronises the ``Period'' sequencing with the switching of 16 NIM logic output lines that can be used for switching external equipment to a defined state at the start of each ``Period''. The outputs are fully programmable via a software interface, making the system highly flexible for carrying out experiment sequences.

\subsection{Other enhancements}
During the design phase of the project, the 0.4 T Helmholtz pair magnet failed. This provided a valuable opportunity to revisit the magnet design and optimise the maximum field given the available cooling water and power supply (2100 A, 100 V). Computer modelling suggested that operation to 0.5 T was possible by forming the magnet from a square section current conductor of side 14 mm with a centre bore for water cooling of 10.5 mm. This was realised in the final system supplied by Sigmaphi\cite{phi}, providing a modest (but scientifically important) increase in the available field range. The field available perpendicular to the muon beam (provided by a pair of water cooled saddle coils) was uprated to 15 mT, while three orthogonal air-cooled coil pairs were reinstated with improved geometry to enable the Earth's magnetic field ($<$0.2 mT) to be compensated for certain experiments.

With a clear requirement to extend the temperature range to dilution temperatures, the instrument vacuum vessel was modified to provide additional space for mounting a facility dilution fridge. The upgraded instrument provides a temperature range from approximately 0.03 K to 1400 K using various liquid helium exchange gas and cold finger cryostats, a closed cycle refrigerator and a reflector furnace based on a design by Major et al~\cite{furnace}.

\section{Commissioning}

Significant time was spent evaluating the performance of the new data acquisition electronics. Tests were devised to confirm the timing resolution, event buffer and to check for artefacts associated with the TDC. Performance of the ``Period'' card was verified by recording sequences of pulse patterns modified by ``Period'' number. In general, tests involved setting up known pulse sequences using a Stanford Research Systems\cite{stanford} DG645 Digital Delay Generator; the unit could be triggered synchronously with the acquisition electronics and the  sequences recorded by the TDC. The artefact test was completed with the DG645 triggered asynchronously with randomly timed pulses recorded by the TDC over an extended period. A Fourier transform of the dataset was carried out to confirm the absence of spurious signals in the TDC output.

A number of calibration measurements were carried out to verify the performance of the finished instrument, with emphasis placed on verifying the performance of the instrument in areas crucial to the success of the project. Performance criteria of particular importance were: the ability to measure at data rates up to $\sim$500 detected events per ISIS frame without requiring deadtime correction for typical datasets, and the ability to make effective deadtime corrections to allow high data rate measurements; stable and fast timing resolution to enable the instrument to be used for pulsed RF measurements, recording signals up to $\sim$100 MHz; a capability for measuring small suspended samples with minimal background signals. Specific examples designed to demonstrate instrument performance in these areas are described in the following sections.

\subsection{Deadtime Corrections}

\begin{figure}[t]
\centering
\includegraphics[width=12cm, trim=1cm 4cm 1cm 1cm]{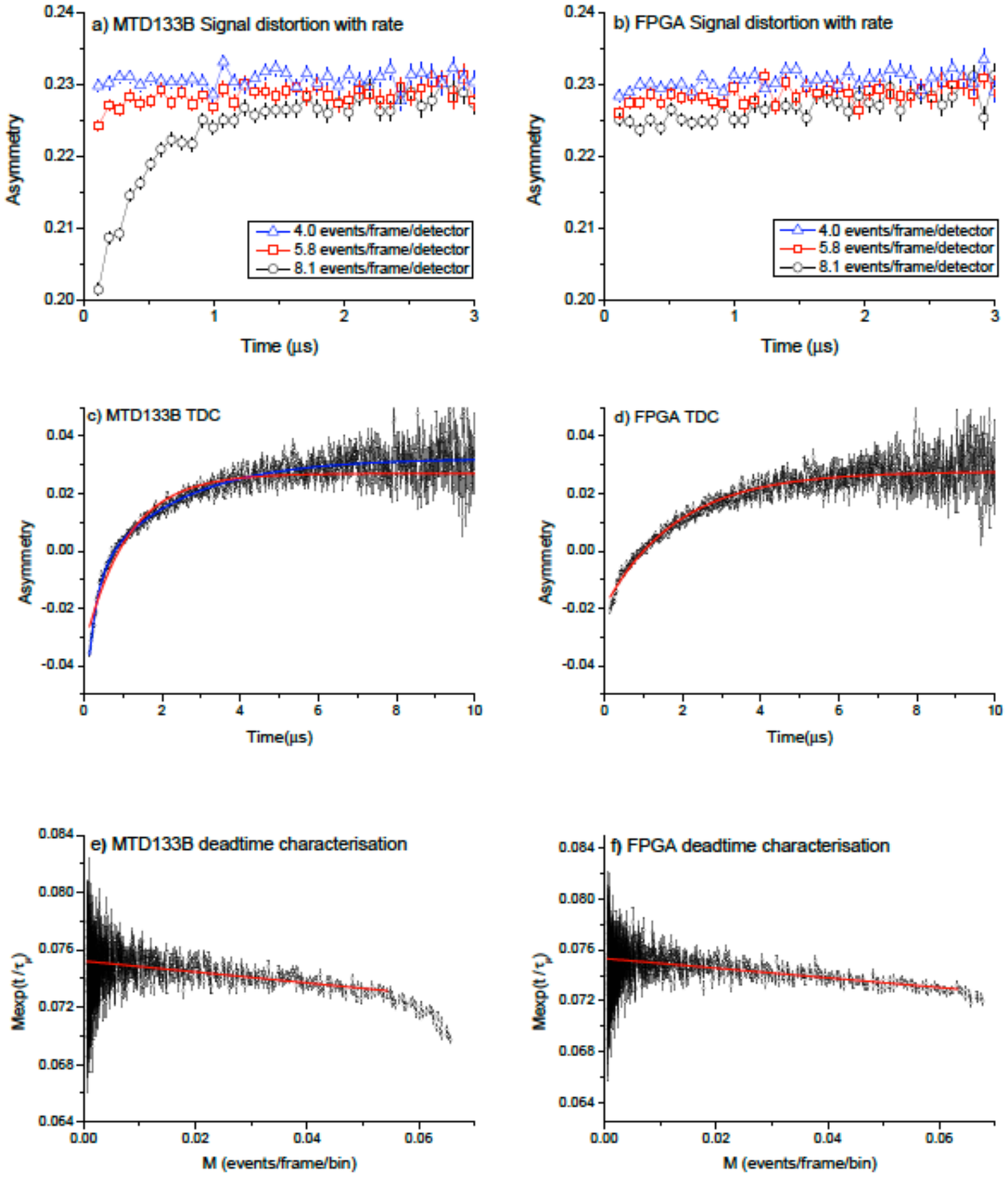}
\caption{a,b) Time dependent plots of the muon asymmetry measured for the full detector array at varying event rates, data being mirrored between the MTD133B and FPGA devices. c,d) Representative data for a single histogram measured with the MTD133B and FPGA devices at $\sim$10.5 events/frame. Fits to the data using a single exponential are shown in red, with the MTD133B requiring a second component to adequately model the signal (shown as a blue line). e,f) Analysis of data measured for the MTD133B and FPGA devices using the non-extendable deadtime model. }\label{deadtime}
\end{figure}

Initial commissioning work investigated the performance of the full detector array by measuring the time dependence of the decay asymmetry for muons stopped in a large high purity silver plate ($> $99.99 $\%$) at room temperature in zero applied magnetic field. Silver was chosen as a suitable target because of the very weak muon spin relaxation in this material owing to small nuclear moments and fast muon diffusion. Care was taken to cancel the Earth's magnetic field at the sample position to avoid slow muon precession complicating the signal. Measurements of $\sim$100x10$^6$ events were made at various counting rates (determined by the setting of the collimation slits) with the data mirrored between the MTD133B and FPGA devices; representative plots are shown in Figure \ref{deadtime}a,b respectively. The difference in the performance of the two TDCs is striking, with early time distortion arising from deadtime effects being seen in the data gathered by the MTD133B device even at 5.8 events/frame/detector, with the data measured at 8.1 events/frame/detector becoming unusable. In contrast, for the FPGA device, only at the highest counting rates does the effect of deadtime become evident, with data measured at 5.8 events/frame/detector being comparable (within error) to low rate measurements and generally usable without further correction.

To investigate the nature of the deadtime distortion for both systems, high statistics measurements ($\sim$1x10$^9$ events) were made at high data rates ($\sim$10.5 events/frame/detector) using an identical silver target. As before, the data was mirrored between the MTD133B and FPGA devices. Individual histograms, each containing $\sim$10x10$^6$ events, are plotted in Figure \ref{deadtime}c,d respectively. The early time distortion of the data for the FPGA device is well represented by a single exponential function, shown by the red curve in Figure \ref{deadtime}d, suggesting the various contributions to the overall deadtime of  the detector segment (scintillator, photomultiplier tube, discriminator and TDC) can be modelled as a single parameter. In contrast, the data obtained for the MTD133B is not well represented by a fit to a single exponential, shown by the red line in Figure \ref{deadtime}c, and a two component exponential fit is required to model the data satisfactorily, shown by the blue line in Figure \ref{deadtime}c. Since all components apart from the TDC are identical to each detector segment, this result suggests the MTD133B is introducing an additional distortion at high data rates that is likely the result of the limited frame buffer. A comparison of the fit parameters supports this conclusion. The time constant for the two components measured for the MTD133B differ by an order of magnitude; however, the slow component is in good agreement (within 15$\%$) to that measured for the FPGA device suggesting a similar source to the deadtime distortion.

A non-extendable deadtime model, following the analysis of Leo~\cite{leo}, can be applied by plotting  $Mexp(t/\tau_\mu)$ against $M$ (where $M$ is the count rate in each histogram bin recorded for each ISIS frame and $\tau_\mu$ the muon lifetime) and should yield a straight line. Any departure of the data from this line suggests the model to be invalid for the measurement system in use. Figure \ref{deadtime}e,f show this analysis carried out for a single histogram recorded by the MTD133B and the FPGA devices respectively; as previously the data is mirrored between the two systems. For the FPGA device, the gradient and intercept of the line provide a composite value for the deadtime for the entire detector chain of $\sim$8 ns. While the data collected by the FPGA device fits the model well over almost the entire range of counting rates, there is a significant departure from the model line for that measured by the MTD133B at the highest data rates. This is expected since at the highest data rates the limited event buffer provided by this device compromises its performance in a way that can't be modelled using this approach. For the FPGA device, representative deadtime values can be obtained for each detector segment in the array and an appropriate correction made in the analysis code to remove early time distortion at high data rates. We have demonstrated that a satisfactory correction can be made even at the  highest data rates currently available on EMU ($\sim$12 events/frame/detector).

\subsection{Timing Resolution probed by Radio Frequency $\mu$SR} 
RF $\mu$SR is an active area of development at the facility~\cite{rf}, with groups using the novel technique to investigate a wide range of topics. One particular advantage of the method is its ability to completely remove the frequency limitations inherent to a pulsed muon source (at ISIS $\sim$6 MHz) due to the finite width of the muon pulse (FWHM $\sim$80ns at ISIS). This is achieved by accumulating the muons for the duration of the muon pulse, with the muon polarisation being preserved in a static magnetic field applied colinear to the muon polarisation. An intense RF pulse is applied immediately following muon implantation, its frequency being chosen to achieve resonance of the muonic species in the static field and the duration adjusted to rotate the muon polarisation by 90$^\circ$. Following the RF pulse, free precession in the static field occurs, the signal decay providing information about the muon environment. The frequency of these free precession signals is generally at least an order of magnitude above the facility pass-band, and therefore this experiment places stringent requirements on the stability and timing resolution of the detector electronics.

Pulsed RF experiments have typically investigated the diamagnetic muon species ($\mu^+$) where the resonance condition is simply determined by the Zeeman energy splitting ($\gamma_{\mu^+}B$). However, the technique is equally applicable to measurement of muons thermalising into the paramagnetic state, muonium. In this case, the resonance condition is determined from the well known Beit-Rabi diagram~\cite{ms} for a one-electron atom with a hyperfine coupling of $\sim$4463 MHz. In a static magnetic field the energy levels associated with the triplet state are split, and measurement of the $\nu_{12}$ and $\nu_{23}$ transitions are readily accessible.

\begin{figure}
\centering
\includegraphics[width=13cm, trim=4cm 0cm 0cm 0cm]{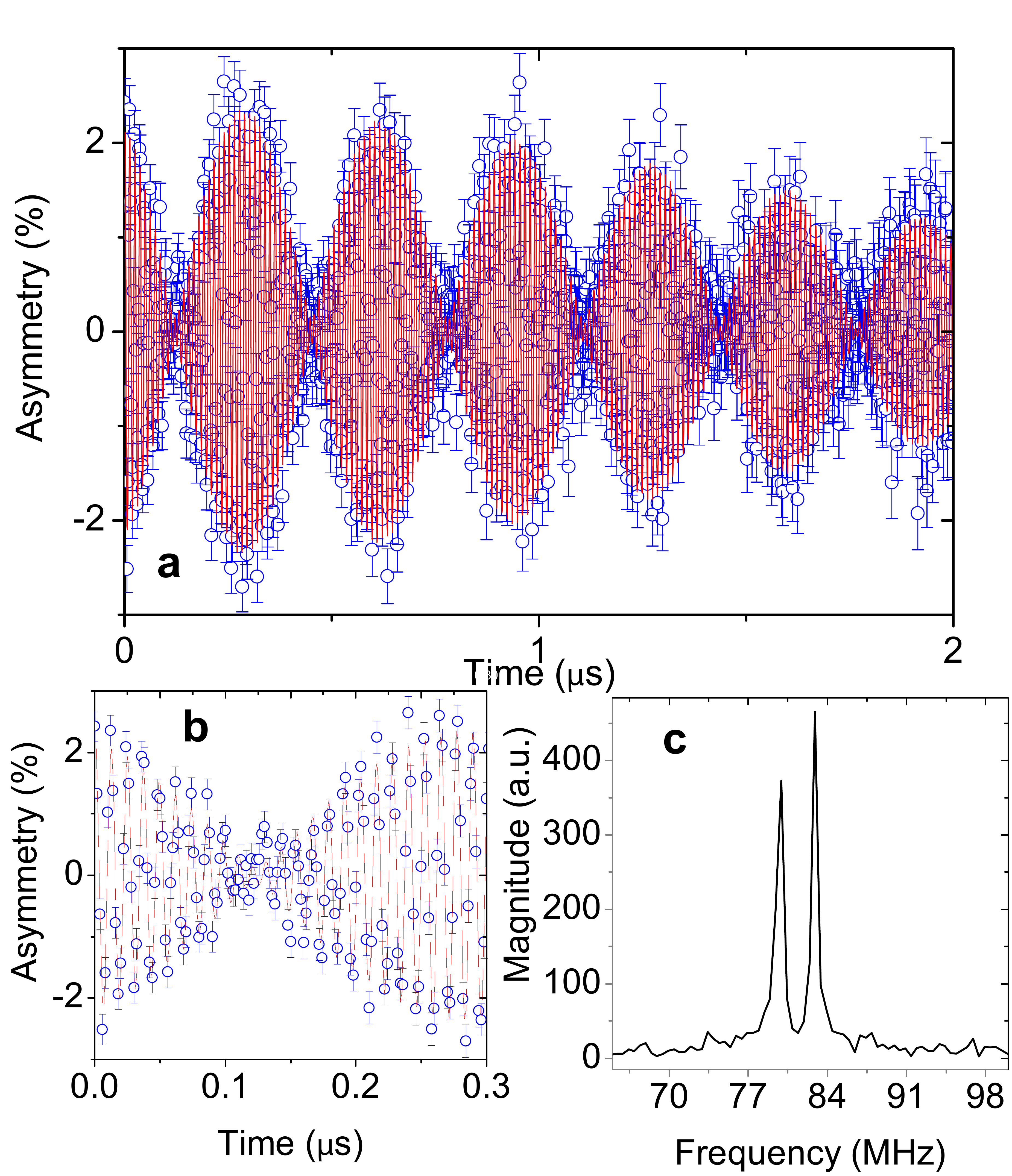}
\caption{a) Free precession signal following a 125 ns 90$^\circ$ RF pulse, both the $\nu_{12}$ and $\nu_{23}$ transitions are excited and recorded in the time domain. b) Expanded portion of the trace showing the TDC timing bin resolution of 2ns. c) The Fourier transform of the data showing two distinct frequencies corresponding to the $\nu_{12}$ and $\nu_{23}$ transitions.}\label{rf}
\end{figure}

For this test, muons were stopped in a quartz target, a material known to have a $\sim$65 $\%$ muonium fraction. Using a static field of 5.8 mT the $\nu_{12}$ and $\nu_{23}$ transitions should occur at 79.4 and 82.3 MHz. The RF field strength was calibrated experimentally and a pulse width of 125 ns determined to provide a 90$^\circ$ rotation of the muon spin polarisation. A particular advantage of the pulsed RF technique is that the short, intense RF pulse excites a band of frequencies centred on the RF carrier ($\sim$80 MHz), and signals associated with both the $\nu_{12}$ and $\nu_{23}$ transitions can be recorded simultaneously in the time domain. This is evidenced by the clear beating pattern measured in the free precession signal shown in Figure \ref{rf}a. An expanded portion of the trace is shown in Figure \ref{rf}b that clearly shows excellent resolution of the 2 ns timing bins used for this measurement (at this resolution the Nyquist frequency of our TDC is 250 MHz, well above the signal frequency). The stability of the TDC was checked by taking a sequence of measurements, initially accumulating 50x10$^6$ events and then progressively doubling the number of events recorded in subsequent measurements. The signal amplitude was checked for each measurement and found, within error, to be identical, providing a good indication that the timing of the TDC is stable over long periods. The data in Figure \ref{rf}a was fitted with two damped cosine functions (shown by the red line in the Figure). The damping of the envelope is either due to impurities in the quartz sample or inhomogeneities in the static magnetic field. A Fourier transform of the entire signal as shown in Figure \ref{rf}c, and clearly shows the two distinct frequencies corresponding to the $\nu_{12}$ and $\nu_{23}$ resonances.

\subsection{Small Samples - Flypast}
The uncollimated muon spot at a pulsed source is comparatively large (at ISIS $\sim$25x10 mm$^2$). While collimation slits can be used to reduce the size of this profile, it is not unusual for experiments to run with $\sim$25$\%$ of the beam falling outside a 20 mm diameter sample area. With a trend towards studying complex high value materials, there is a clear need to be able to obtain good results when investigating small samples ($<$10 mm diameter). The new instrument was designed with an enhanced capability for measuring suspended samples, an established technique~\cite{lynch} for measuring small samples at a pulsed muon source, where muons not implanted in the sample continue in an evacuated flight tube and are removed from the sample position. Rate reductions of approximately a factor of three are typical for measurements of a 5x5 mm$^2$ suspended sample when compared to that of a large plate; however, this is readily compensated by the enhanced sample asymmetry obtained using the technique.

To evaluate the performance of the spectrometer in this area we used Polytetrafluoroethylene (PTFE) and silver (Ag) as test samples due to the unique and well defined muon response that can be used both to quantify the ratio of sample signal to background and to investigate any distortion arising from stray counts. Plates of size 40x40 mm$^2$ were used for measurements stopping all incident muons, while suspended samples of size 5x5 mm$^2$ were mounted on a 0.5 mm thick silver blade orientated edge-on to the incident beam. Commissioning tests were carried out in the dilution fridge cryostat - with the straight-through beam passing through a total of six windows this represents a particularly challenging environment in which to obtain clean data. Measurements were made with a low pressure helium exchange gas at 50 K to ensure implanted muons were static in the sample material.

In high purity Ag a Gaussian relaxation of the muon signal is expected that can be modelled using equation~\ref{gaussian}. Comparatively weak nuclear moments result in only a small depolarisation of the muon signal in zero magnetic field, with $\Delta\approx$ 0.02 MHz.
\begin{equation} \label{gaussian}
A(t)=A_0exp(- \Delta ^2t^2/2)
\end{equation}
In PTFE the muon is known to form a hydrogen-like bond with two fluorine atoms, the system is commonly known as an F-$\mu$-F state. The spectrum for this species is a unique oscillation and is a direct consequence of the strong muon-fluorine dipole-dipole coupling. The original  observation of the F-$\mu$-F state was made by Brewer et al~\cite{fl} during a study of muon spin relaxation in the Alkali fluorides. The relaxation function, $G_{\mathrm{F}\mu \mathrm{F}}$, can be solved from first principles assuming a three spin model (two fluorine atoms and one muon) to give $G_{\mathrm{F}\mu \mathrm{F}}$ as a function of time ($t$) of the form:
\begin{equation} \label{FuF}
\resizebox{.92\hsize}{!}
 {$G_{\mathrm{F}\mu \mathrm{F}} = \frac {1}{6} \left [3 + cos(\sqrt{3}\omega_d t) + 
\left ( 1- \frac{1}{\sqrt{3}} \right ) cos \left(  \frac{3- \sqrt{3}}{2} \omega_d t \right ) + 
\left ( 1+ \frac{1}{\sqrt{3}} \right ) cos \left(  \frac{3+ \sqrt{3}}{2} \omega_d t \right ) \right ] $}
\end{equation}
where $\omega_d = \frac{\mu_0 \gamma_{\mu} \gamma_{F}}{4\pi} \left \langle \frac{1}{r^3} \right \rangle$, $\gamma_{\mu}$ is the muon gyromagnetic ratio, $\gamma_{F}$ is the $^{19}$F gyromagnetic ratio, r is the $\mu$-F bond length which is averaged over the bond lengths which result from motion of the polymer chain. Therefore the experimental relaxation can be modelled using the functional form:
\begin{equation} \label{muonFuF}
A(t)=A_1G_{\mathrm{F}\mu \mathrm{F}}exp(- \lambda t)+A_2exp(- \Delta ^2t^2/2)+A_{bg}
\end{equation} 
where $A_{1}$ is the amplitude of the F-$\mu$-F state, $A_{2}$ the amplitude associated with muons not forming the F-$\mu$-F state which experience a quasi-static distribution of fields ($\Delta$) due to the $^{19}$F nuclear moment, and $A_{bg}$ describing a signal background. The relaxation, $\lambda$, originates from the dynamics of the F-$\mu$-F state.

\begin{figure}
\centering
\includegraphics[width=14cm]{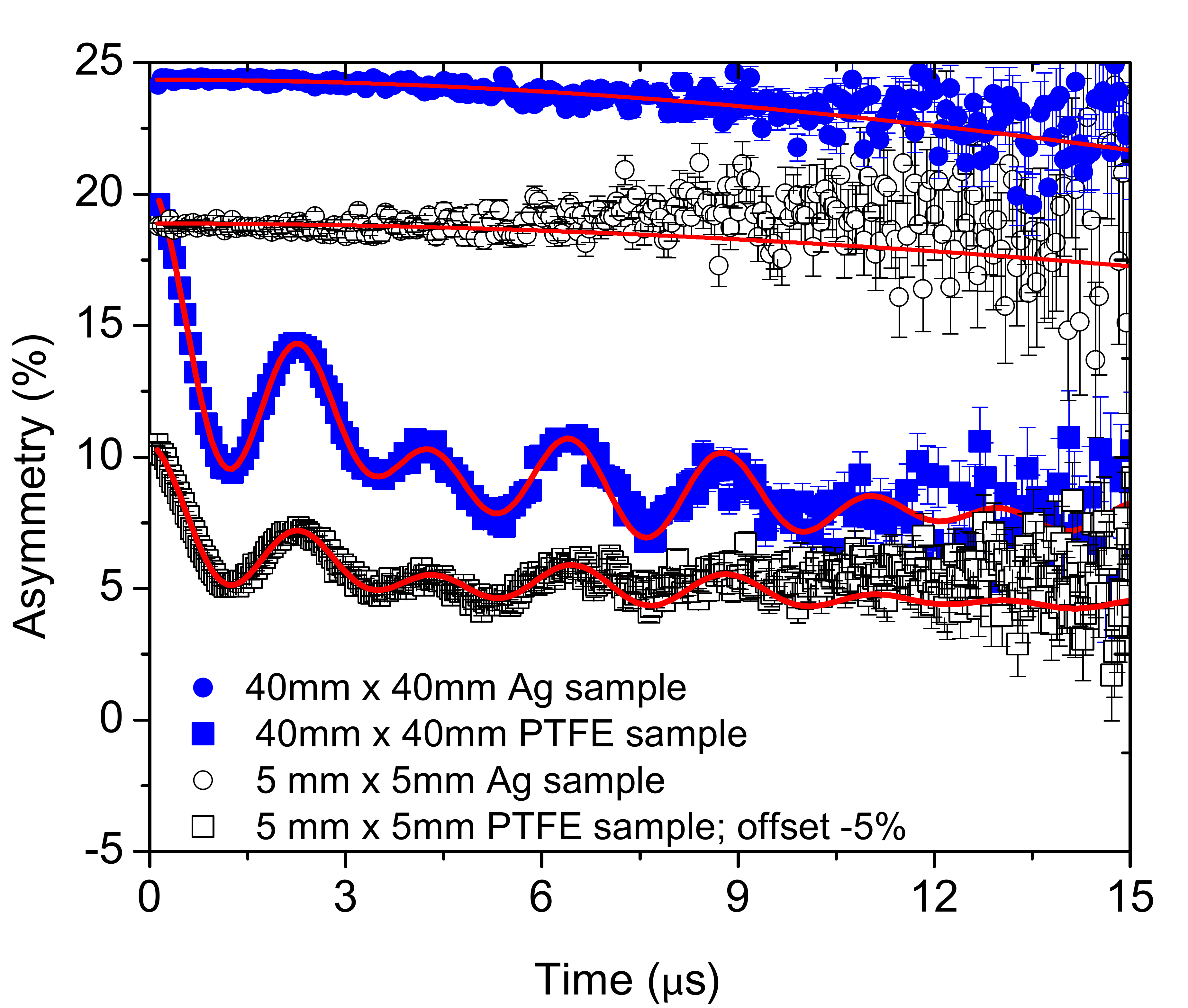}
\caption{Time dependence of the Ag and F-$\mu$-F asymmetry, showing oscillations from the F-$\mu$-F state formed in PTFE and the Gaussian relaxation in the Ag for both a large plate and a small suspended sample. All measurements at 50K.}\label{ptfe}
\end{figure}

The zero field signal for muons stopped in a large Ag plate ($\sim$90x10$^6$ events) is shown in Fig. \ref{ptfe}; the data are fitted by Equation \ref{gaussian} to give $\Delta = 0.022\pm0.001$ MHz. Comparative data was measured for the 5x5 mm$^2$ Ag suspended sample (28x10$^6$ events) and results are also shown in Fig. \ref{ptfe}. Immediately apparent is the reduction of the measured asymmetry, which is probably the result of muons offset from the optimal sample position contributing counts to both the forward and backward detector banks - aside from modifying the overall signal amplitude this appears to do little to alter the fitted parameters. A similar fit to the first 5 $\mu$s of the data using Equation \ref{gaussian} yields a value for $\Delta$ of $0.024\pm0.005$ MHz, in good agreement to results obtained for the large plate. In this case, however, there is an apparent distortion of the signal at long times, most likely due to background contributions from counts arising from environmental sources. This ÔnoiseÕ can become significant when measuring suspended samples because the relatively low muon flux on the sample requires extended counting times for adequate statistics. However, variable temperature measurements following the onset of muon diffusion in Ag demonstrate that very small changes in the muon relaxation can readily be measured for the 5x5 mm$^2$ suspended sample.

To study the relative contribution of the sample signal to the overall measured asymmetry it was necessary to select a sample material with a signal clearly distinct from likely background contributions. PTFE, with a clear F-$\mu$-F signal, is an ideal choice for this calibration. A high statistics measurement for muons stopped in a large PTFE plate is shown in  Figure \ref{ptfe}; a clear signature of the F-$\mu$-F state is seen. In this case, a fit to the data using Equation \ref{muonFuF} gives values for $A_{1} = 11.8 \pm0.1\%$ and $A_{2} = 1.70 \pm0.1\%$, essentially determining the population of each muon state in the material. These results can be compared to similar measurements for a 5x5 mm$^2$ PTFE suspended sample; as expected the total asymmetry is reduced, with Equation \ref{muonFuF} giving values of $A_{1} = 6.6 \pm0.1\%$ and $A_{2} = 0.9 \pm0.1\%$. Taking the asymmetry of the F-$\mu$-F state, $A_{1}$, as a proportional measure of the total sample asymmetry, this result suggests that the asymmetry for the suspended sample is $\sim$56\% of that measured for the large plate. A commensurate reduction in the total asymmetry was not seen; however, it is noted that fits for the data measured for the suspended sample show a significant increase in the non-relaxing background signal (modelled by $A_{bg}$). This is presumably due to muons scattered around the cryostat entry windows (masked by silver plate) and stopping in the silver sample support. Crucially, the data for the suspended sample is clean without artefacts, which could arise from muons stopping in the flight tube giving rise to an additional background signal source. The relaxation rates and observed frequencies are also, to within error, identical to those measured for the large plate.

\section{Conclusions}
We have developed a state-of-the-art muon spectrometer for the ISIS pulsed muon source as a major upgrade of the highly successful EMU instrument original commissioned at the facility in 1993. In tandem with enhanced data acquisition electronics, the instrument is able to make effective use of the full muon flux currently available at the facility. With an appropriate correction for detector deadtime, facility users are able to measure at detected event rates of up to 140x10$^6$ events/hour assuming an uncollimated beam. These counting rates are of huge benefit, enhancing the quality of typical datasets while making possible the development of novel pulsed muon techniques that exploit the pulsed structure of the ISIS beam - an area of experimental spectroscopy ripe for exploration. During commissioning work we have demonstrated that detector deadtime can be quantified and corrected for each detector segment, that the new data acquisition has a frequency resolution and stability necessary for working with pulsed RF experiments and that the instrument is capable of working with small samples, with a signal quality enhanced with the redesigned spectrometer. The new instrument is now playing a full role in the ISIS muon user programme.

\end{document}